\documentstyle[12pt]{article}
\begin{document}
\begin{titlepage}
\normalsize
\begin{center}
{\Large \bf Budker Institute of Nuclear Physics} 
\end{center}
\begin{flushright}
BINP 96-48\\
July 1996
\end{flushright}
\vspace{0.5cm}
\begin{center}
{\Large\bf Nature of the Darwin term} 
\end{center}
\begin{center}
{\Large\bf and ${(Z\alpha)^4 m^3/M^2}$ contribution to the Lamb shift}
\end{center}
\begin{center}
{\Large\bf for an arbitrary spin of the nucleus}
\end{center}

\vspace{0.5cm}

\begin{center}
{\bf I.B. Khriplovich}\footnote{e-mail address: khriplovich@inp.nsk.su},
{\bf A.I. Milstein}\footnote{e-mail address: milstein@inp.nsk.su}
\end{center}
\begin{center}
Budker Institute of Nuclear Physics, 630090 Novosibirsk, Russia
\end{center}

\begin{center}
{\bf and R.A. Sen'kov}
\end{center}

\begin{center}
Novosibirk University
\end{center}

\vspace{0.5cm}

\begin{abstract}
The contact Darwin term is demonstrated to be of the same origin as
the spin-orbit interaction. The $(Z\alpha)^4 m^3/M^2$ correction to
the Lamb shift, generated by the Darwin term, is found for an
arbitrary nonvanishing spin of the nucleus, both half-integer and
integer.  There is also a contribution of the same nature to the
nuclear quadrupole moment.
\end{abstract}

\end{titlepage}

{\bf 1.} The literature, pedagogical included, abounds with
assertions on the nature of the Darwin correction which are at least
doubtful in our opinion. In particular, we cannot agree with the
conclusion that the Darwin term is absent for a particle with spin
$1$, made in Ref.\cite{r4} (see also \cite{r5}). The subject becomes
of real interest now for interpreting the high precision experiments
in atomic spectroscopy \cite{r1,r2,r6}.

To study the problem we consider in this note the Born amplitude for
scattering of a particle with an arbitrary spin in an external
electromagnetic field.  In the case of a practical interest, that of
an atom, this is the nucleus interaction with the electromagnetic
field of electron. In this way we derive the general form of the
Darwin term for an arbitrary nuclear spin and obtain the
corresponding order $(Z\alpha)^4 m^3/M^2$ correction to the Lamb
shift (here and below $Z$ and $M$ are, respectively, the charge and
mass of the nucleus).

\bigskip

{\bf 2.}  The wave function of a particle with an arbitrary spin can
be written as (see, for instance, \cite{r7}, \S 31)
\begin{equation}\label{ps} 
            \Psi=\left( \begin{array}{c} \xi\\
                                         \eta\\ 
                        \end{array} 
                 \right).  
\end{equation} 
Both spinors,
\[
\xi=\{ \xi^{{\alpha}_1\,{\alpha}_2\,\,..\,\,{\alpha}_p\,}_{\dot{\beta}_1
\,\dot{\beta}_2\,..\,\,\dot{\beta}_q}\} \]
and
\[ \eta=\{ \eta_{\dot{\alpha}_1\,\dot{\alpha}_2
\,\,..\,\,\dot{\alpha}_p\,}^{{\beta}_1\,{\beta}_2\,..\,\,{\beta}_q}\},
\]
are symmetric in dotted and undotted indices separately, and
\[ p+q=2I, \] 
where $I$ is the particle spin. In the rest frame $\xi$ and $\eta$ 
coincide and are symmetric in all indices. For a particle of
half-integer spin one can choose
\[ p=I+\,\frac{1}{2}\;\;,\;\;q=I-\,\frac{1}{2}\,\,. \] 
In the case of integer spin it is convenient to take
\[ p=q=I. \] 
Spinors $\,\xi\,$ and $\,\eta\,$ are choosen in such a way
that under reflection they go over into each other (up to a phase).
At $p\,\neq\,q$ they are different objects which belong to different
representations of the Lorentz group. If $p=q$, these two spinors
coincide. Nevertheless, we will use the same expression ($\ref{ps}$)
for the wave function of any spin, i.e., we will introduce formally
the object $\,\eta\,$ for an integer spin as well, keeping in mind of
course that it is expressed via $\,\xi\,$. It will allow us to
perform calculations in the same way for both integer and
half-integer spins.

The Lorentz transformation from the rest frame is, up to to the terms
$\sim (v/c)^2$ included, 
\begin{eqnarray}
\xi=\left(1+\frac{\vec{\Sigma}\vec{v}}{2}+
\frac{(\vec{\Sigma}\vec{v}\,)^2}{8}\right)\xi_0\,,\nonumber\\
\eta=\left(1-\frac{\vec{\Sigma}\vec{v}}{2}+
\frac{(\vec{\Sigma}\vec{v}\,)^2}{8}\right)\xi_0\,.  \end{eqnarray}
Here
\[ \vec{\Sigma}\,=\,\sum_{i=1}^{p} \vec{\sigma}_i\,-\,
\sum_{i=p+1}^{p+q} \vec{\sigma}_i\,, \]
and ${\vec{\sigma}}_i$ acts on the $i$th index of the spinor $\,\xi_0\,$
as follows:
\begin{equation}\label{aa}
\vec{\sigma}_i\,\xi_0
=(\vec{\sigma}_i)_{\alpha_i\beta_i}\, (\xi_0)_{....\beta_i...}\;.
\end{equation}
By analogy with the spin $1/2$, let us introduce, in line with the
``spinor" representation (\ref{ps}), the ``standard" one:
\[ \phi=(\xi+\eta)/2\,;\;\;\chi=(\xi-\eta)/2\,. \]
In it the wave function is written as
  \begin{equation}
   \Psi=\left(
               \begin{array}{c}
                 (1+(\vec{\Sigma}\vec{v}\,)^2/8)\,\xi_0\\
                    {\vec{\Sigma}\vec{v}}/2\,\xi_0\\
               \end{array}  
          \right)\,.     
  \end{equation}
Let us note that
\[ \bar{\Psi}\,\Psi=
  \phi^{\star}\phi-\chi^{\star}\chi=\xi_0^{\star}\,\xi_0 \]
is an invariant. We will use however the common noncovariant
normalization of the particle number density
  \begin{equation}
     \rho=\frac{E}{M}\bar{\psi}\psi=1,
  \end{equation}
where the wave function $\psi$ is
  \begin{equation}
      \psi=\sqrt{\frac{M}{E}}
          \left(\begin{array}{c}
                 \left(1+{(\vec{\Sigma}\vec{v}\,)^2}/8\right)\,\,\xi_0\\
                   {\vec{\Sigma}\vec{v}}/2\,\,\xi_0\\
               \end{array}  
           \right).     
  \end{equation}

\bigskip
 
{\bf 3.} Let us go over now to the scattering amplitude itself. The
order $\,1/M^2\,$ terms in it arise only in the time component of the
electromagnetic current.  Restricting to the formfactors of the
lowest multipolarity, electric $F_e$ and magnetic $G_m$, this
component can be written for an arbitrary spin as
\begin{equation}\label{rho}
j_0={F}_1\frac{E+E^{\prime}}{2M}\bar{\psi}^{\prime}F_e\psi
    +\frac{{G}_m}{2M}\psi^{\prime\,\star}\,\vec{\Gamma}\vec{q}\,\psi.
\end{equation}  

The matrix
  \begin{equation}\label{aaa}
      \vec{\Gamma}=\left(
                            \begin{array}{rr}
                                0 & {\vec{\Sigma}}\\
                               {-\vec{\Sigma}} & 0\\ 
                             \end{array}
                       \right)    
      \end{equation}
is a natural generalization of the corresponding expression for spin
$1/2$ (valid both in the spinor and standard representations):
\begin{equation}\label{1/2}
 \vec{\gamma}=\left(
                           \begin{array}{rr}
                               0 & \vec{\sigma}\\
                              -\vec{\sigma} & 0\\                 
                           \end{array}      
                     \right)\;.  
\end{equation}                     
This generalization is fairly obvious in the spinor representation.
Indeed, here, according to (\ref{1/2}), $\vec \sigma$ connects a
dotted index in the initial spinor $\psi$ with undotted one in
$\bar\psi$, and $\;-\,\vec \sigma$ connects an undotted index from
$\psi$ with dotted one of $\bar\psi$. And this is exactly what is
being done by $\vec\Gamma$.  It is straightforward now to prove the
expression (\ref{aaa}) for the standard representation. Let us
mention also that formula (\ref{aaa}) is confirmed by the final
result which reproduces correctly the spin-orbit interaction, the
form of the latter being well-known for an arbitrary spin (see, e.g.,
\cite{r7}, \S 41).
 
The term with $G_m$ in the current density is
     \begin{eqnarray}
       j_{0\,m}=\frac{G_m}{2M}\xi_0^{\prime\,\star} 
         \left(\begin{array}{cc}
                 1, & \vec{\Sigma}\vec{v}\,^{\prime}/2\\
                \end{array}
         \right)
                      \left(\begin{array}{rr}
                                0 & {\vec{\Sigma}}\vec q\\
                               -\vec{\Sigma}\vec q & 0\\
                            \end{array}
                      \right)
                               \left(\begin{array}{c}
                                        1\\
                                        \vec{\Sigma}\vec{v}/2\\
                                     \end{array}
                               \right)\xi_0\nonumber\\
        =\,\frac{G_m}{4M^2}\xi_0^{\prime\,\star}
\left(-(\vec{\Sigma}\vec q)^2\,+\,4\,i\,\vec{I}[\vec{q}\times\vec{p}\,]
\right)\,\xi_0\;.
\end{eqnarray}
The spin operator here equals
\[ \vec{I}\,=\,\frac{1}{2}\,\sum_{i=1}^{2I}\vec{\sigma_i}\,. \]
The first term, with $F_e$, in formula (\ref{rho}) reduces to an
analogous structure:
\begin{eqnarray}
j_{0\,ch}=F_e\frac{E^{\prime}+E}{2\sqrt{EE^{\prime}}}
         \,\xi_0^{\prime\,\star}
                  \left(
                      1+\frac{(\vec{\Sigma}\vec{v})^2}{8}
                       +\frac{(\vec{\Sigma}\vec{v}^{\prime})^2}{8}
                       -\frac{(\vec{\Sigma}\vec{v}^{\prime})
                                (\vec{\Sigma}\vec{v})}{4}
\right)\xi_0\nonumber\\
         =F_e\,\xi_0^{\prime\,\star}
                    \left(
                         1+\frac{(\vec{\Sigma}\vec q)^2}{8M^2}                          
          -\,i\,\frac{\vec{I}[\vec{q}\times\vec{p}\,]}{2M^2}  
\right)\xi_0.
\end{eqnarray}
Thus the total charge density is
\[ j_0=\xi_0^{\prime\,\star} \left(
F_e - (2G_m-F_e)\frac{(\vec{\Sigma}\vec{q}\,)^2}{8M^2}
+(2G_m-F_e)\,i\,\frac{\vec{I}[\vec{q}\times\vec{p}\,]}{2M^2} \right)\xi_0.
  \]
We neglect for the time being the charge radius of the nucleus, so
that
\[ F_e=F_e(0)=1. \]
The spin-orbit interaction dependence on the gyromagnetic ratio $g$
is universal for any spin, this ratio enters through the factor
$g-1$. Therefore, our magnetic formfactor is normalized as follows
\[ G_m(0)=\frac{g}{2}. \]

Let us split now $(\vec{\Sigma}\vec q)^2$ into the contact and
quadrupole parts:
     \begin{equation}\label{kk}
          \Sigma_i\Sigma_j\,q_iq_j\,=\,\frac{\vec q\,^2}{3}\Sigma_i\Sigma_i
          +(q_iq_j-\frac{1}{3}\vec q\,^2\delta_{ij})\Sigma_i\Sigma_j\\ 
     \end{equation}
The first, contact term in (\ref{kk}) is
  \begin{eqnarray}\label{zeta}
\vec{\Sigma}\vec{\Sigma}=\left(\,\sum_{i=1}^p\vec{\sigma}_i\right)^2
       -2\left(\,\sum_{i=1}^p\vec{\sigma}_i \right)
 \left(\sum_{i=p+1}^{p+q}\vec{\sigma}_i \right)
+\left(\,\sum_{i=p+1}^{p+q}\vec{\sigma_i} \right)^2\nonumber\\
       = 3(p+q)+2\left( \frac{p(p-1)}{2}+\frac{q(q-1)}{2}-pq
                 \right)\,=\,4I(1+\zeta);
   \end{eqnarray}
  \[ \zeta=\left\{
                   \begin{array}{cc}
                              0, & {\rm integer spin,}\\
                              1/(4I), & {\rm halfinteger spin.}\\ 
           \end{array} \right. \]
In derivation of formula (\ref{zeta}) we use the symmetry in any pair
of spinor indices, $\alpha_1\;\alpha_2$ (see (\ref{aa})). This
symmetry means that the corresponding spins, 1 and 2, add up into the
total spin $S=1$. Therefore, 
 \[ (\vec{\sigma}_1\vec{\sigma}_2)\, \xi_0=\xi_0\,. \]

The interaction operator is proportional to the Fourier transform of
the Born amplitude (see, e.g., \cite{r7}, \S 83). In this way we
obtain from (\ref{zeta}) the following contact interaction between a
nucleus of charge $Z$ and electron:
\begin{equation}
      U(\vec{r})=\frac{2\,\pi}{3}\,\frac{Z\alpha}{M^2}\,(g-1)\,I\,(1+\zeta)\,
                    \delta(\vec{r})\,.
\end{equation}                         
The corresponding energy correction is
\begin{equation}\label{de}
      \Delta E_n=\frac{2}{3}\frac{m^3}{M^2}\,
                (g-1)\,I\,(1+\zeta)\,\frac{(Z\alpha)^4}{n^3}\delta_{0\ell}. 
  \end{equation} 
For the hydrogen atom ($I=1/2$) this correction was obtained long ago
in Ref. \cite{r3}.
   
Let us consider now the quadrupole part of (\ref{kk}). Using again
the complete symmetry of $\xi_0$, one can easily calculate the
corresponding quadrupole interaction:
   \begin{equation}\label{u2}
U_2(\vec r)=-\frac{1}{6}\,\nabla_i\nabla_j\frac{e}{r}\,\delta Q_{ij}\,.
   \end{equation}
Here
   \begin{equation}\label{q}
        \delta Q_{i\,j}=\,-\,\frac{3}{4}\,\frac{Z\,e\,(g-1)}{M^2}\,\Lambda\,
           \left\{\,I_i\,I_j+I_j\,I_i-\frac{2}{3}\delta_{i\,j}\,I(I+1)\,
           \right\}\,;
   \end{equation}  
\[ \Lambda=\left\{
                   \begin{array}{cc}
                              1/(2I-1), & {\rm integer spin\,,}\\
                              1/(2I), & {\rm halfinteger spin\,.}\\
                   \end{array}
             \right. \]
Expression (\ref{q}) is a correction to the nuclear quadrupole
moment. Its existence for $I=1$ was pointed out in Ref. \cite{r4}.

This correction to the quadrupole moment can be estimated as
\[
\delta Q\approx\,\,-\,0.22\,(g-1)\,\frac{Z\,I}{A^2}\;
e\,\mbox{mbarn}\,.
\]
For the deuteron ($Z=1,\;A=2,\;g=2\mu_d=1.714,\;Q=2.86\;e\,$mbarn) 
it equals $\;-0.04\;e\,$mbarn.

\bigskip
   
{\bf 4.} Let us come back now to the discussion of the contact term.
There is some ambiguity in its definition related to the nuclear
charge radius. The contribution of the latter produces a contact
interaction also and enters physical observables in a sum with the
expression $(g-1)\vec q^2\;I(1+\zeta)/(6M^2).$ In particular, the elastic
cross-section of the electon-nucleus scattering at small $\vec q\,^2$
is, up to the terms $\vec q\,^2/M^2$  included,
\[ \frac{d\sigma}{d\Omega}\,=\,\frac{\alpha^2}{4\epsilon^2}\,
\frac{\cos^2\theta/2}{\sin^4\theta/2}\,
\frac{1}{1+\,2\,\sin^2\theta/2\,\epsilon/M}\, \]
\begin{equation}\label{sca} 
     \left( [1-\frac{1}{6}\,\langle r^2 \rangle_F\,\vec q\,^2
                 -(g-1)\,\frac{\vec q\,^2}{6M^2}I(1+\zeta)]^2 \right.
\end{equation}               
\[ \left. +\,\frac{4}{3}\,G_m^2\,I(I+1)(2\,\tan^2\theta/2+1)\right), \]
where $\langle r^2 \rangle_F$ is defined through the expansion of the
formfactor $F_e$:
\begin{equation} 
F_e(q^2)\,\approx\,1-\,\frac{1}{6}\,\langle r^2 \rangle_F\,\vec{q}\,^2.  
\end{equation} 
Let us note here that the expression in square brackets in formula
(\ref{sca}) reduces for the proton ($I=1/2$) to
\begin{equation}\label{pro}
1-\,\frac{1}{6}\,\langle r^2 \rangle_F\,\vec q\,^2
                 -(g-1)\,\frac{\vec q\,^2}{8 M^2},                                        
\end{equation}
and for the deuteron ($I=1$) to
\begin{equation}\label{deu}
1-\,\frac{1}{6}\,\langle r^2 \rangle_F\,\vec q\,^2
                 -(g-1)\,\frac{\vec q\,^2}{6 M^2}.                                        
\end{equation}

However, the proton charge radius is commonly defined otherwise than
in formula (\ref{pro}), namely, through the expansion of the
so-called Sachs formfactor
\[ G_e=F_e\,-\,\frac{\vec q\,^2}{4M^2}\,G_m\,. \]
Obviously, the charge radius defined through the formfactor $G_e$ is
\[ -\frac{1}{6}\langle r^2 \rangle_G\, =\, \frac{\partial G_e}{\partial
\vec q\,^2}\, =\,-\,\frac{1}{6}\,\langle r^2 \rangle_F\, 
-\,\frac{g}{8 M^2}. \] 
Correspondingly, expression (\ref{pro}) is rewritten usually as
\begin{equation}\label{pro1}
1-\,\frac{1}{6}\,\langle r^2 \rangle_G\, \vec q\,^2
+\,\frac{\vec q\,^2}{8 M^2},                                        
\end{equation}
and the Darwin correction for the proton is defined as
\[ \frac{\vec q\,^2}{8 M^2}\,, \]
but not 
\[ -\,\frac{(g-1)\vec q\,^2}{8 M^2}\,. \]

We could redefine the electric formfactor for the deuteron from $F_e$
to $G_e$ in such a way that here
\[ -\,\frac{1}{6}\,\langle r^2 \rangle_G\, 
=\, \frac{\partial G_e}{\partial \vec q\,^2}\,
=\,-\,\frac{1}{6}\,\langle r^2 \rangle_F\, -\,\frac{g}{6 M^2}, \]                                      
so that the Darwin correction for the deuteron becomes
\[ \frac{\vec q\,^2}{6 M^2}\,, \]
instead of 
\[ -\,\frac{(g-1)\vec q\,^2}{6 M^2}\,. \]
However, for a deuteron the common definition of the charge radius is
neither $F_e$, nor $G_e$, but
\[ -\frac{1}{6}\langle r^2 \rangle_D\, 
=\,-\,\frac{1}{6}\,\langle r^2 \rangle_F\, -\,\frac{g-1}{6 M^2}. \]                                      
Of course, under this definition the whole Darwin term is swallowed
up by $\;\langle r^2 \rangle_D\,$. No wonder therefore that the
authors of Ref. \cite{r4}, using $\;\langle r^2 \rangle_D\,$ instead
of $\;\langle r^2 \rangle_E\,$ or $\;\langle r^2 \rangle_G\,,$ make
the conclusion that for the deuteron, as distinct from proton, the
Darwin correction is absent.

Clearly, this contradistinction of the deuteron to proton is based
only on a rather arbitrary definition of the charge radius of the
former; this contradistinction has no physical meaning, it has
nothing to do with the nature of the Darwin term.

\bigskip

{\bf 5.} Thus, the Darwin interaction exists for any nonvanishing
spin and is of the same nature as the spin-orbit interaction. In
particular, as well as the spin-orbit interaction, the Darwin term is
not directly related to the so-called Zitterbewegung. Of course,
there is a certain difference between the spin-orbit and contact
energy corrections. The former one has a classical limit together
with $\langle 1/r^3 \rangle$, while the latter, being proportional to
$|\psi(0)|^2$, does not. However, this fact has nothing to do with
relativity and negative energies, and therefore is certainly
unrelated to the Zitterbewegung.

\end{document}